\documentclass[rmp,twocolumn,showpacs]{revtex4}

\usepackage{graphicx,amsmath,amssymb,txfonts,dcolumn}

\begin{document}

\title{Exploring the assortativity-clustering space of a network's
  degree sequence}

\author{Petter Holme}
\affiliation{Department of Computer Science, University of New Mexico,
  Albuquerque, NM 87131, U.S.A.}

\author{Jing Zhao}
\affiliation{School of Life Sciences \& Technology, Shanghai Jiao Tong
  University, Shanghai 200240, China}
\affiliation{Shanghai Center for Bioinformation and Technology,
  Shanghai 200235, China}
\affiliation{Department of Mathematics, Logistical Engineering
  University, Chongqing 400016, China}

\begin{abstract}
  Nowadays there is a multitude of measures designed to
  capture different aspects of network structure. To be able to say if
  the structure of certain network is expected or not, one needs a
  reference model (null model). One frequently used null model is the
  ensemble of graphs with the same set of degrees as the original
  network. In this paper we argue that this ensemble can be more than just
  a null model---it also carries information about the original network and
  factors that affect its evolution. By mapping out this ensemble in the
  space of some low-level network structure---in our case those
  measured by the assortativity and clustering coefficients---one can
  for example study how close to the valid region of the parameter
  space the observed networks are. Such analysis suggests which
  quantities are actively optimized during the evolution of the
  network. We use four very different biological networks to exemplify
  our method. Among other things, we find that high clustering might
  be a force in the evolution of protein interaction networks. We also
  find that all four networks are conspicuously robust to both random
  errors and targeted attacks.
\end{abstract}

\pacs{89.75.Fb, 82.39.Rt, 89.75.Hc}

\maketitle

\section{Introduction}

Network structure~\cite{mejn:rev,doromen:book,ba:rev} is usually
defined as the way a network differs from what is expected. What
``expected'' means depends on the fundamental constraints on the
network, and this can vary from system to system. For example, if
the network is made of units that must be connected to two, and only
two, others; then, it is not interesting whether or not a vertex lies
on a cycle (we already know that it will). The ensemble of all
networks fulfilling the fundamental constraints on the system is
usually called \textit{null model} (or \textit{reference model}). When
we have pinned down the null model we can measure the network
structure by standard quantities. If the values of these quantities
differs significantly from the null-model average, then we call the network
structured. The baseline assumption of complex network theory is that
network structure carries information about the forces that have
formed the network. Ever since the studies of Barab\'{a}si and
coworkers~\cite{ba:model,ba:rev}, the degree distribution (or, if
referring to the set of degrees of one particular network,
\textit{degree sequence}) has been regarded as the most fundamental
network structure. For many networks, the degrees are related to outer
factors (not emerging from the network evolution). In such cases the
ensemble of all graphs with the same degree sequence as the original
network is a natural null model. Another interpretation is that the
network structures measured relative to this null model are of higher
order than the degree---i.e., what remain after the effects of the
more fundamental structure (the degree sequence) is filtered away. The
usual way to use a null model is to compare a network measure with the
ensemble average value of the null model. In this paper we will argue
that one can glean more information about the original network by studying
the null model ensemble in greater detail than just measuring
averages.

We consider networks that can be modeled as a graph $G=(V,E)$ where
$V$ is the set of $N$ vertices and $E$ is the set of $M$ undirected
edges. We denote the ensemble of graphs with the same degree sequence
as $G$ as $\mathcal{G}(G)$. Our basic approach to study
$\mathcal{G}(G)$ is to resolve its members in the space of higher
order network structures. The two such higher order network structures
we consider in this paper are: the correlation between the
degrees at either side of an edge (measured by the \textit{assortative
  mixing coefficient}, $r$~\cite{mejn:assmix}, or simply
\textit{assortativity}); and, the fraction of triangles in the network
(measured by the \textit{clustering
  coefficient}, $C$~\cite{bw:sw,mejn:rev}). By mapping out
$\mathcal{G}(G)$ in the space defined by $r$ and $C$ one can pose
questions such as: How large is the region in $r$-$C$ space where
members of $\mathcal{G}(G)$ actually exist? (This helps us answer how
constrained the network evolution is if the degrees are given.) Is the
real network close to $\mathcal{G}(G)$'s boundaries in $r$-$C$ space?
(Which would indicate whether or not $r$ or $C$ are actively
optimized.)

The basis for our exploration of an ensemble $\mathcal{G}(G)$ is to map
out its members in the space defined by some network-structural
measures, in our case the assortativity and clustering. We explore the
$r$-$C$ space by successively rewire pairs of edges, $(i,j)$ and
$(i',j')$ to $(i,j')$ and $(i',j)$, that takes the system in a desired
direction. Rewiring techniques for studying networks are half a century
old~\cite{gale:rew} (randomization for obtaining null models was
studied in Ref.~\cite{katz:cug}). In the physics literature these
techniques were first used in Refs.~\cite{maslov:pro,alon}.

\section{Network structural measures}

Before going into details of our algorithm, we will review the network
structural quantities that we use to describe our networks: both the
independent variables (the assortative and clustering
coefficients) that form the basis for our space of interest; and the
quantities we use for characterizing the regions of this space.

\subsection{Assortative mixing coefficient}

It is quite well accepted that the set of degrees, the degree
sequence, is the network quantity that contains most information about both the evolution and function of the network. Degree can (in most
contexts) be identified as how influential the vertex is~\cite{wf} (in
some sense)---high
degree vertices are assumed to be more influential both the formation
of the network and the flow of dynamic systems on the network. In this
paper we assume the degree sequence is inherent to the system and look
at higher order structures arising from how the vertices are linked to
one another. The simplest such higher-order structure is the
correlations between the degrees of vertices at either side of an
edge. Is it the case that high-degree vertices are primarily connected
to other high degree vertices, or are they linked to low-degree
vertices? A simple way of measuring this tendency is by the
assortativity~\cite{mejn:rev} $r$. Basically
speaking, $r$ is the linear correlation coefficient of the degrees at
either side of an edge. One complication is that since the edges are undirected, $r$ has to be symmetric with respect to edge-reversal, but the correlation coefficient is not symmetric. The solution is to let one edge contribute
twice to the covariance, i.e.\ represent an undirected edge by two directed edges pointing in opposite directions. If one use an edge list representation
internally (i.e., let the edges be stored in an array of ordered pairs
$(i_1,j_1),\cdots,(i_M,j_M)$) then~\cite{mejn:assmix}
\begin{equation}\label{eq:assmix}
  r=\frac{4\langle k_1\, k_2\rangle - \langle k_1 + k_2\rangle^2}
  {2\langle k_1^2+k_2^2\rangle - \langle k_1+ k_2\rangle^2}
\end{equation}
where, for an edge $(i,j)$, $k_1$ is the degree of first argument
(i.e., the degree of $i$) and $k_2$ is the degree of the second
argument. The range of $r$ is $[-1,1]$ where negative values indicate
a preference for high connected vertices to attach to low-degree
vertices, and positive values means that vertices tend to be attached
to others with degrees of similar magnitudes.

\subsection{Clustering coefficient}

Several simple random network models (such as the
Edr\H{o}s-R\'{e}nyi~\cite{er:on} or the model for generating networks
of a given $r$-value in Ref.~\cite{mejn:assmix}) have rather few triangles (fully connected
subgraphs of three vertices). For some classes of real-world networks (notably
social networks~\cite{holl:72}) there is a strong tendency for triangles to form, which makes such models fail. The network measure of the density of
triangles is called \textit{clustering coefficient}. We use the definition
of Ref.~\cite{bw:sw}:
\begin{equation}\label{eq:clust}
  C = 3 n_\mathrm{triangle}\:\big/\:n_\mathrm{triple},
\end{equation}
where $n_\mathrm{triangle}$ is the number of triangles and
$n_\mathrm{triple}$ is the number of connected triples (subgraphs
consisting of three vertices and two or three edges). The factor three
is included to normalize the quantity to the interval $[0,1]$.

\subsection{Distance and component size}

Two quantities that are, perhaps more than any other, related to the functionality of dynamic processes on the network are the relative size of the largest
component (connected subgraph) $s$, and the average distance $\langle
d\rangle$. $s$ is simply defined as the number of vertices in the
largest component divided by $N$. The distance $d(i,j)$ between two
vertices $i$ and $j$ is defined as the number of edges in the shortest
path between these two vertices. $\langle d\rangle$ is $d(i,j)$
averaged over all vertex pairs ($i\neq j$) in the largest
component. In a network with large $s$ and small $\langle d\rangle$,
spreading processes will be fast and far-reaching. This is a good
property of information networks, but bad in the context of, for
example, disease spreading. Some authors have combined the distance
and component size aspects by considering the average reciprocal
distances~\cite{our:attack,latora:eff}. For most purposes, we believe,
valuable information gets lost in such a combination (a fragmented
network $G$ with short average distances can be something very
different from a connected graph of large distances and the same
average reciprocal distances as $G$).

\subsection{Robustness}

One line of complex network research is the study of the response of
the network to attacks, errors, failures and other events that effectively change
the structure. The error response problem is usually formulated as: how
does the functionality of the network change if a random fraction of
the vertices, or edges, is removed~\cite{mejn:rev}? The attack
problem is the same, except that the vertices are not selected
randomly but according to some strategy intended to decrease the
networks' functionality as rapidly as possible~\cite{our:attack,alb:attack}.
A frequently used metric for functionality is the ratio of $s$
before and after the
event~\cite{our:attack,alb:attack,motter:cascade}. In the error and
attack robustness problems, this quantity is typically plotted as a
function of the number of removed vertices. The idea is that even if
one network $G$ is more robust than another network $G'$ to the
removal of, say, $1\%$ of the vertices, $G'$ can be less vulnerable
than $G$ if $10\%$ of the vertices are deleted. Since we aim at
mapping out the $r$-$C$ space of degree sequences, we would like to
capture the robustness with just one number. We will use what we call
the $f$-\textit{robustness} $R_f$ of a network as the expected
fraction of vertices that needs to be removed for the relative size of
the largest component to decrease to a fraction $f\in (0,1)$ of its
original value. The way of removal can either be random (the error
problem) or selective (the attack problem). For the rest of the paper
we will set $f=1/2$, and refer to the $1/2$-robustness just as
``robustness'' $R$. Other $f$-values give slightly different results, but
our conclusions will hold for a range of intermediate $f$-values.

 \begin{figure}
  \centering\resizebox*{0.9\linewidth}{!}{\includegraphics{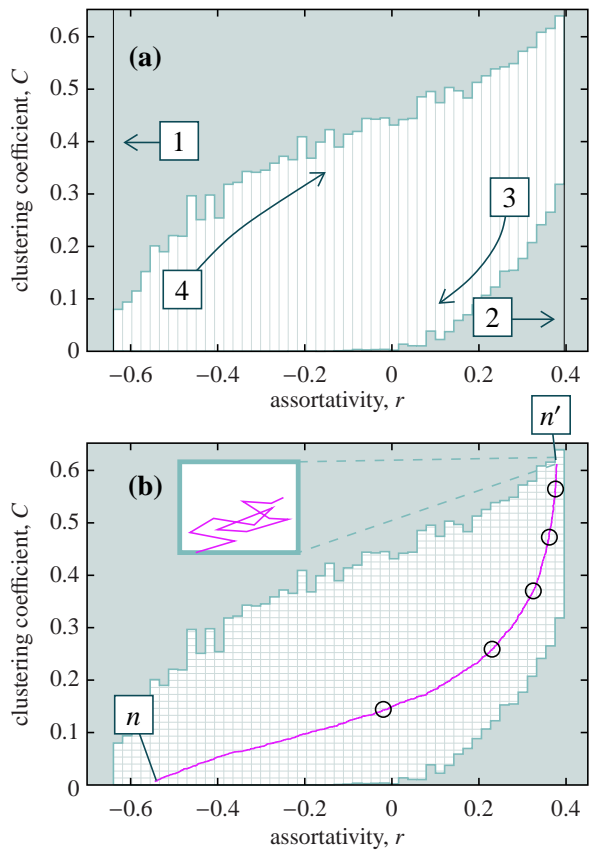}}
  \caption{Illustration of the analysis scheme applied to the
    \textit{C. elegans} neural network. (a) shows how the valid region
    is mapped out: 1. $r_\mathrm{min}$ is located. 2. $r_\mathrm{max}$
    is found and the interval $[r_\mathrm{min}, r_\mathrm{max}]$ is
    divided into $L$ segments. 3. $C_\mathrm{min}(n)$ is
    constructed. 4. $C_\mathrm{max}(n)$ is traced and the interval
    $[C_\mathrm{min}, C_\mathrm{max}]$ is segmented into $L$
    regions. (b) illustrates the sampling of the pixels. The next
    pixel to go to is chosen from a random permutation of the
    pixels. In this example $n$ and $n'$ are chosen to be far
    apart. The line shows the path taken by the algorithm. The circles
    indicate every thousandth step on the way from $n$ to $n'$. The
    blow-up illustrates the random walk within a pixel to sample the
    graphs of the pixel more randomly.
}
  \label{fig:ill}
\end{figure}

\section{The analysis scheme\label{sect:analysis}}

The fundamental idea of our method is simple: we update the network by
choosing pairs of edges randomly, say $(i,j)$ and $(i',j')$, and swap
one end of them (forming $(i,j')$ and $(i',j)$). This guarantees that
the degree sequence stays intact. We navigate in the $r$-$C$ space by
only accepting changes that move us in the desired direction.  If an
edge-swap would introduce a self-edge (i.e.\ if $i=j'$ or $i'=j$) or a
multiple edge (i.e.\ if $(i,j')$ or $(i',j)$ belongs to $E$ before the
swapping, or \textit{move}) it is not performed. There are many other
technicalities concerning the convergence to extremes, uniformity of
the sampling and more that we discuss in the Appendix.

The members of the ensemble $\mathcal{G}(G)$ do not, in general, cover
the whole range of $(r,C)$-values. Indeed, for any finite $G$,
$\mathcal{G}(G)$ defines a set of points, rather than a continuous
region, in the $r$-$C$ space. We will perform a more coarse-grained
analysis breaking down the $r$-$C$ space into pixels and average quantities over the graphs of $\mathcal{G}(G)$ with $(r,C)$-values within the pixel. (Thus, a pixel constitute a graph ensemble in itself, our aim is to sample its members with uniform randomness.) For a computationally tractable
resolution, the pixels containing members of $\mathcal{G}(G)$
typically form contiguous regions. We will refer to the pixels that
contain a member of $\mathcal{G}(G)$ as \textit{valid pixels}, and all
pixels that are valid or between valid pixels the \textit{valid
  region} of $\mathcal{G}(G)$.

To trace the valid region of $\mathcal{G}(G)$ we start by finding the
lowest and highest assortativity value, $r_\mathrm{min}$ and
$r_\mathrm{max}$ respectively. Briefly speaking (more details follow
below), to find $r_\mathrm{min}$ we rewire edge-pairs that lower $r$
(and vice versa for $r_\mathrm{max}$). After finding the extremal
$r$-values, we splice the region between these into $L$ segments. Then
we go through the region and for each region $n\in [1,L]$ we find the
minimal and maximal $C$-values, $C_\mathrm{min}(n)$ and
$C_\mathrm{max}(n)$. The region in $C$-space between the lowest
$C_\mathrm{min}=\min_{1\leq n\leq L}C_\mathrm{min}(n)$ and highest
$C_\mathrm{max}=\max_{1\leq n\leq L}C_\mathrm{max}(n)$ observed
clustering coefficient is segmented into $L$ regions. (Note that $C_\mathrm{min}$, without argument,
is the global clustering minimum, whereas $C_\mathrm{min}(n)$ is the
minimum conditioned on $r$ being in the $n$'th segment.) Thus we
(assuming our method works) obtain an $L\times L$ grid of the $r$-$C$
space that contains the valid region of $\mathcal{G}(G)$. The method
is illustrated in Fig.~\ref{fig:ill}.

To find the $\mathcal{G}(G)$ elements of minimal and maximal
assortativity is a non-trivial optimization problem. There are
deterministic methods that, if they terminate, are guaranteed to give the
maximal (or minimal) assortativity~\cite{doyle:big,zj:spectrum}. To avoid the
such technicalities and to simplify the program, we will use the same
kind of optimization algorithm to find $r_\mathrm{max}$ and
$r_\mathrm{min}$ as to find $C_\mathrm{min}(n)$ and
$C_\mathrm{max}(n)$. In the Appendix we will argue that this method
allows us to come as close to the optimal $r$-values as we need.
A method we find efficient is to repeat the
simple edge-pair swapping procedure (where only changes in the desired
direction are accepted) with different random seeds until no lower
state is found during a number $\nu_\mathrm{rep.}$ of
repetitions~\cite{walker_walstedt}. Each individual edge-pair is
terminated when no lowest state is found for $\nu_\mathrm{same}$
swaps. In general, the larger the network is, the more densely
distributed are the points close to the border of the valid region. If
one is satisfied with finding a value a certain distance from the
extrema, then $\nu_\mathrm{rep.}$ and $\nu_\mathrm{same}$ do not need to be
increased for larger $N$. To find $C_\mathrm{min}(n)$ and
$C_\mathrm{max}(n)$ almost the same procedure is employed. First,
edge-pairs are swapped until the desired segment of $r$ is
found. Second, unless $r$ is outside the segment $n$ and the move
takes the system yet further from segment, edge-pairs are swapped provided
the clustering would decrease (for $C_\mathrm{min}(n)$), or increase,
(for $C_\mathrm{max}(n)$). When the valid region is traced out and we
sample networks of different pixels, we select the pixels
randomly. The idea is to sample the space of networks more randomly.

To summarize, the algorithm for finding the extremal assortativity
values, $r_\mathrm{min}$ and $r_\mathrm{max}$, is:
\begin{enumerate}
\item \label{step:choose} Choose two undirected edges $(i,j)$ and
  $(i',j')$ at random. If the program makes a difference between the
  arguments of the edge, the direction of the reading of the edge also
  has to be randomized (so $(i,j)$ is read as $(j,i)$ with probability
  $1/2$).
\item \label{step:check} Check if swapping these edges to $(i,j')$ and
  $(i',j)$ would introduce a self-edge or multiple edge in the
  network. If so, go to step~\ref{step:choose}.
\item \label{step:accept} Let $\Delta r$ be the change in $r$ if the
  move in step~\ref{step:choose} is executed. If $r$ is
  to be minimized and $\Delta r<0$, then accept the change (vice versa for maximization of $r$).
\item \label{step:conclude} If no move has been executed during the last
  $\nu_\mathrm{same}$ executions of step~\ref{step:accept}, then take
  the current $r$ as $\tilde{r}_\mathrm{min}$ (or $\tilde{r}_\mathrm{max}$).
\item \label{step:stop} Repeat from the beginning $\nu_\mathrm{rep.}$
  times and return the lowest observed $\tilde{r}_\mathrm{min}$ during these iterations.
\end{enumerate}

Given $r_\mathrm{min}$ and $r_\mathrm{max}$, and a division of the $r$
space into $L$ segments of width $(r_\mathrm{max}-r_\mathrm{min})/L$,
we trace the boundaries of the valid region as follows:
\begin{enumerate}\setcounter{enumi}{5}
\item \label{step:choose2} Go through the regions sequentially. Say
  the $n$'th region is the interval $[r_n,
  r_{n+1})$.
\item Perform step~\ref{step:choose} and \ref{step:check} of the
  assortativity optimization algorithm.
\item Let $\Delta C$ be the change in clustering coefficient during the previous
  step. If $r<r_n$ and $\Delta r>0$, $r\geq
  r_{n+1}$ and $\Delta r<0$ or $r_n \leq r <
  r_{n+1}$ and $\Delta C < 0$ (for minimization) or $\Delta
  C > 0$ (for maximization), then perform the change of
  step~\ref{step:choose2}.
\item \label{step:conclude2} If, counting from the first time the
  system entered the desired $r$-segment, the minimal (maximal)
  $C$-value has been repeated $\nu_\mathrm{same}$ times, take this
  value as $\tilde{C}_\mathrm{min}(n)$ ($\tilde{C}_\mathrm{max}(n)$).
\item \label{step:stop2} Repeat from step~\ref{step:choose2}
  $\nu_\mathrm{rep.}$ times. Let the lowest
  $\tilde{C}_\mathrm{min}(n)$-values and largest
  $\tilde{C}_\mathrm{max}(n)$ during these iterations be
  $C_\mathrm{min}(n)$ and $C_\mathrm{max}(n)$.
\end{enumerate}

Then, when the valid region is mapped out, we split the $C$-range
(between $C_\mathrm{min}$ and $C_\mathrm{max}$ in $L$ segments of
equal width, thus forming an $L\times L$-grid enclosing the valid
region. This grid is sampled as follows:
\begin{enumerate}\setcounter{enumi}{10}
\item \label{step:perm} Construct a random permutation of the valid
  pixels.
\item \label{step:pick} Pick the next pixel
  $P_n=[r_n,r_{n+1})\times
  [C_m,C_{m+1})$ from the index-list of
  step~\ref{step:perm}. Denote the center $[(r_n+r_{n+1})/2,(C_m+C_{m+1})/2)]$ of the pixel
  $(r_{n,0},C_{m,0})$. Let
  \begin{equation}\label{eq:dist}
    \delta (r,C)=\sqrt{\left(\frac{r - r_{n,0}}{r_\mathrm{max} -
          r_\mathrm{min}}\right)^2 + \left(\frac{C -
          C_{m,0}}{C_\mathrm{max}-C_\mathrm{min}}\right)^2}
  \end{equation}
  measure the distance in $r$-$C$ space from the current position
  $(r,C)$ to the center of the target pixel.
\item Pick edge-pair candidates according to steps~\ref{step:choose}
  and \ref{step:check} of the assortativity optimization algorithm.
\item Calculate $\Delta (r,C)=\delta(r',C')-\delta(r,C)$ where $r$
  and $C$ are the current assortativity and clustering values, and
  $r'$ and $C'$ are the values after the pending move has been
  performed. If $\Delta (r,C)<0$ perform the move.
\item \label{step:rw} If the updated $(r,C)$ belongs to $P_n$, then: First, make
  $\nu_\mathrm{rnd.}$ random edge swappings such that $(r,C)$ does not
  leave $P_n$. (This is to sample the pixel more uniformly.) Then,
  measure network structural quantities of $P_n$, save these values
  for statistics, and go to step~\ref{step:pick}.
\item If not all pixels have been measured go to step~\ref{step:pick}.
\item Go to step~\ref{step:perm} until each pixel have been sampled
  $\nu_\mathrm{samp.}$ times.
\end{enumerate}

The parameter values we use in this study are (unless otherwise stated):
$\nu_\mathrm{same}=10^5$, $\nu_\mathrm{rep.}=5$,
$\nu_\mathrm{samp.}=100$, $\nu_\mathrm{rnd.}=1000$ and $L=50$. The
choice of parameters and further considerations are discussed in the
Appendix. Due to the uncertain stopping conditions of
steps~\ref{step:conclude}, \ref{step:stop}, \ref{step:conclude2} and
\ref{step:stop2} it is hard to derive meaningful bounds on the
computational complexity. We note, however, that the optimization is
faster in $r$- than in $C$-direction, this probably relates to the observation in
Fig.~\ref{fig:ill}(b) that swapping procedure moves faster in the $r$-
than in the $C$-direction. (The speed in the $C$-direction is roughly the
same per 1000 steps, but the speed in the $r$-direction decrease.) 

\section{Networks}

Our method can be applied to every kind of system that can be modeled
as an undirected network.  To limit ourselves, we use four networks
from biology as examples in this paper. These networks are,
nonetheless, representing fundamentally different systems.

\begin{table}
\caption{Basic statistical properties of the example networks we
  use. The number of vertices $N$, number of edges $M$,
  assortativity $r$, clustering coefficient $C$, relative size of the
  largest cluster $s$, average distance in the largest cluster
  $\langle d\rangle$, the error robustness $R_\mathrm{error}$ and the
  attack robustness $R_\mathrm{attack}$.}
\begin{ruledtabular}
  \begin{tabular}{r|dddd}\label{tab:stat}
    & \multicolumn{1}{c}{gene fusion} &
    \multicolumn{1}{c}{protein interaction} &
    \multicolumn{1}{c}{metabolic} &
    \multicolumn{1}{c}{neural}\\\hline
    $N$ & 291 & 4168 & 1905 & 280 \\
    $M$ & 278 & 7434 & 3526 & 1973 \\
    $r$ & -0.36 & -0.13 & -0.10 & -0.069 \\
    $C$ & 0.0016 & 0.034 & 0.039 & 0.20 \\
    $s$ & 0.38 & 0.94 & 0.87 & 1 \\
    $\langle d\rangle$ & 4.2 & 4.8 & 4.5 & 2.6 \\
    $R_\mathrm{error}$ & 0.43 & 0.36 & 0.36 & 0.50 \\
    $R_\mathrm{attack}$ & 0.012 & 0.048 & 0.046 & 0.38 \\
  \end{tabular}
\end{ruledtabular}
\end{table}

\subsection{Gene fusion network}

Cancer is a disease that occurs due to changes in the genome. One
important process causing such changes is gene fusion---when two
genes merge to form a hybrid gene~\cite{mitelman}. In
Ref.~\cite{hoglund} the authors construct a network of human genes
that have been observed to be fused in the development of tumors in
humans. Some genes can fuse with many others but most of the genes have
only been observed fusing with one, or a few others. The resulting
network structure has a skewed, power-law like degree distribution and
is rather fragmented---the largest component spanning only $38\%$ of the
vertices. Statistics of this and the other networks are listed in
Table~\ref{tab:stat}.

\subsection{Metabolic network\label{sect:meta}}

A cell can be regarded as a machine driven by biochemical reactions. The
possible reactions of the metabolism (the cellular
biochemistry except signaling processes) and its environment
determine the state of the cell. The metabolism of an organism is a very
complex system---so complex that one has to choose between studying
a part of it in detail, or the whole with a coarser method. One approach in the latter category is to construct a network, connecting the chemical
substrates occurring in the same reactions to a network, and employ
network analysis to characterize the large-scale structure of the
metabolism. The way to construct a biochemical network is not entirely
straightforward~\cite{zhao:meta}. Should the substances be
linked to each other (in a \textit{substrate graph}), or to the
reactions they participate in? If one use a substrate graph, should
the substrates be linked only to products, or to all reactants
(i.e.\ in a reaction A + B $\leftrightarrow$ C + D, should A be
linked to C and D, or to all three other vertices)? Furthermore, some
chemical substances (like H$_2$O, ATP, NADH, and so on) are abundant
throughout the cell and seldom pose any restriction on the reaction
dynamics. For many purposes, one obtains a more meaningful network by
deleting such \textit{currency metabolites}. The biochemical network
we use is the human metabolic network of Ref.~\cite{our:curr}. In this
network, substrates are linked only to products (A to C and D in the
above example). Currency metabolites are identified and deleted
according to a self-consistent, graph-theoretic method~\cite{our:curr}.

\subsection{Protein interaction}

In protein interaction networks the vertices are proteins and two
proteins constitute an edge if they can interact physically. Examples
of interaction are the ability to form complexes, carrying another
protein across a membrane or modifying another protein. We use the
(``physical interaction'') data set from Ref.~\cite{hh:pfp}
of protein interaction in the budding yeast \textit{S. cerevisiae}.

\subsection{Neural network}

For the biochemistry of an organism, the network representation is a
crude model of the system as a whole (as an alternative to a detailed
model of a subsystem). Neural networks are yet more complex. For these
the choices are either to make a coarse-grained network
representation~\cite{sporns:cortex} or study the full network of a very
simple organism. In this work, we take the latter approach and study
the neural network of \textit{C. elegans}~\cite{white:420}. In this
data set, the strength of the neuronal coupling has been measured, but we make the network undirected by
letting an edge represent a non-zero coupling.

\begin{figure*}
  \centering\resizebox*{0.75\linewidth}{!}{\includegraphics{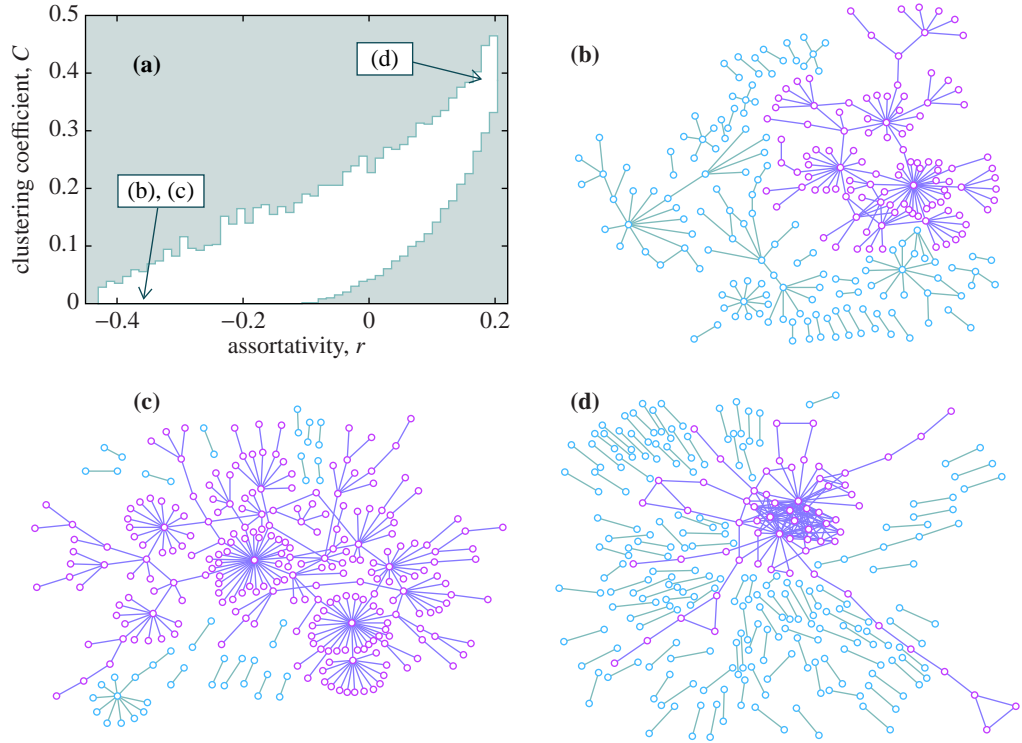}}
  \caption{The valid region demarcated by the $C_\mathrm{min}(r)$- and
    $C_\mathrm{max}(r)$-curves (a), and three networks: (b) is the
    original gene fusion network, (c) shows a random sample with
    $r$-$C$ coordinates close to those of the real network. (d) shows a network
    with high clustering and high assortativity. The largest component
    of (b), (c) and (d) are indicated with a different color.
  }
  \label{fig:exp}
\end{figure*}

\section{Numerical results}

In this section we present numerical results for our four
network-structural measures over the $\mathcal{G}(G)$ ensembles of the
four test graphs. To get a first view, we display the valid region of the
gene fusion graph in Fig.~\ref{fig:exp}(a). As seen, the valid region is
not covering a large part of the theoretical limits of $r$ ($-1\leq r\leq 1$) and
$C$ ($0\leq C < 1$). (Note that only fully
connected graphs have $C=1$, and for these $r$ is undefined.) The
requirement that the graph should be simple (no multiple edges or
self-edges) puts hard constraints on the actual $r$-values that can
occur (cf.\ Ref.~\cite{maslov:inet}). Fig.~\ref{fig:exp}(a) shows that, considering the entire $r$-$C$ plane, such constraints are even harder.
The general shape of the valid region is consistent with
the observations that the simple-graph constraint induce a positive
correlation between $r$ and $C$~\cite{maslov:inet,mejn:why}.

In Fig.~\ref{fig:exp}(b), (c) and (d) we show three example networks
of $\mathcal{G}(G)$ (where $G$ is the gene fusion
network). Fig.~\ref{fig:exp}(b) displays the relatively fragmented
real network. Fig.~\ref{fig:exp}(c) is a random network $G'$ with the
almost the same $r$-$C$ coordinates as the real network
($\delta(G,G')\approx 0.0026$). Maybe the biggest visible difference
between $G$ and $G'$ is the larger size of the largest component of
$G'$. Is it true that the gene fusion network is unusually fragmented,
given the degree sequence and $r$-$C$ coordinates? If so, there might
be an evolutionary pressure for gene fusion networks to be fragmented. (This
will be discussed further in Sect.~\ref{sect:size}.)
Fig.~\ref{fig:exp}(d) shows, as a contrast, a network far away from
$G$ and $G'$. The network has a well-defined core where high-degree
vertices connect to each other. There are also a number of peripheral
triangles, which indicates that the network evolves toward a maximal
$C$-value, given its assortativity.

\begin{figure*}
  \centering\resizebox*{0.75\linewidth}{!}{\includegraphics{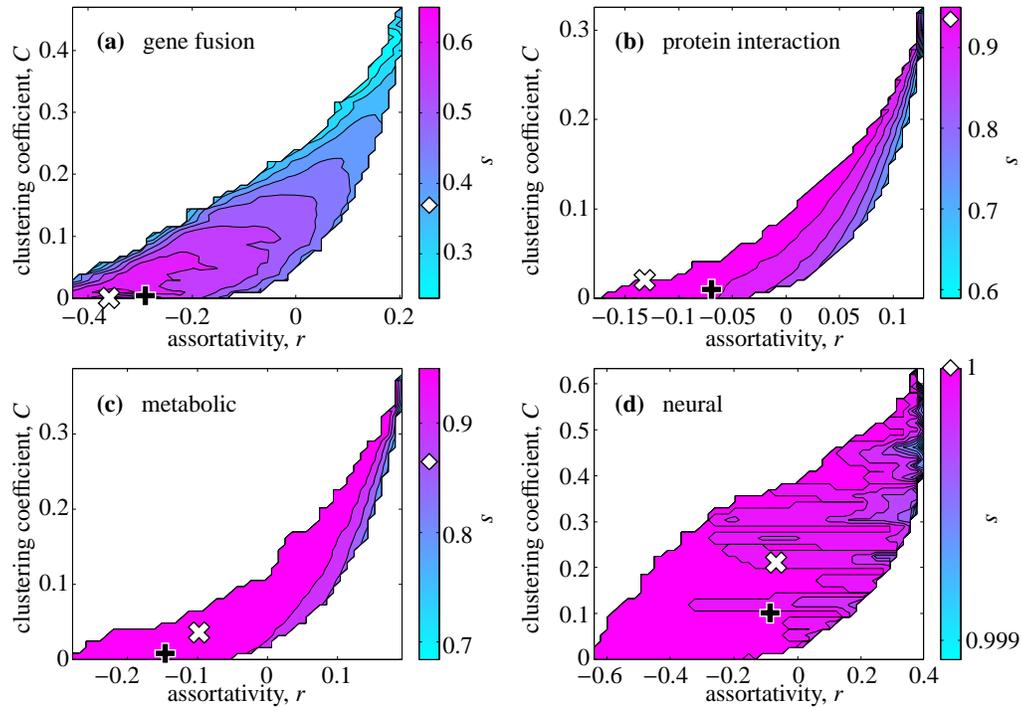}}
  \caption{The relative size of the largest component $s$ as a function of
    $r$ and $C$. The networks are (a) the network of gene fusions in
    tumors in humans, (b) protein interaction network of
    \textit{S. cerevisiae}, (c) human metabolic network and (d) the
    \textit{C. elegans} neural network. The x-like symbols of the main
    figures and the diamond symbols of the color-bars indicate
    the values of the real networks. The plus-like symbol
    indicates the average $(r,C)$-value of the $\mathcal{G}(G)$
    ensemble.}
  \label{fig:ngc}
\end{figure*}

\subsection{Location in $r$-$C$ space and size of largest
  component\label{sect:size}}

In Fig.~\ref{fig:ngc} we plot the relative size of the largest
component of the four test networks. We also display the locations of
the actual networks in the $r$-$C$ plane, and the
$\mathcal{G}(G)$-averages. (The $\mathcal{G}(G)$ averages are obtained
from a rewiring sampling of $\mathcal{G}(G)$, with
step~\ref{step:check} of the algorithm as the only constraint.) We see
that the $C$-value of the gene fusion graph lies close to the
$C_\mathrm{min}(r)$-boundary of its valid region. $C$ averaged over
the whole $\mathcal{G}(G)$ is about three times larger ($\langle
C\rangle_{\mathcal{G}(G)}=0.0061\pm 0.0001$) than the observed value
($C=0.0017$). Furthermore, we see that the assortativity is lower than
the $\mathcal{G}(G)$ average. This kind of analysis has been used by
many authors (following Ref.~\cite{maslov:pro}). The interpretation is
usually that the network is, effectively, disassortative and clustered
(i.e., $r<\langle r\rangle_{\mathcal{G}(G)}$ and $C>\langle
C\rangle_{\mathcal{G}(G)}$). However, looking at the entire valid
region, we can get another perspective: If high clustering really
would have been an important goal for the network to obtain (given the
degree sequence) there is large room for improvement. For the
assortativity, on the other hand, the observed network is rather close to the
minimum. This might be telling us that assortativity is a more
important factor, than clustering, in the evolution of the gene fusion
networks. The protein interaction network of Fig.~\ref{fig:ngc}(b) is
located quite far from the ensemble average---the assortativity is much
lower than the $\mathcal{G}(G)$-average, and given that assortativity,
the clustering is maximal. Also
the metabolic (Fig.~\ref{fig:ngc}(c)) and neural
(Fig.~\ref{fig:ngc}(d)) networks are more clustered than the average,
but here the assortativity is slightly larger than the
$\mathcal{G}(G)$ average. From Fig.~\ref{fig:ngc} we also note that
the density of states is very inhomogeneous distributed---the
average $(r,C)$ is close to $C=0$ and (except for the neural network)
left of the middle of the assortativity spectrum. The shapes of the
valid regions are rather similar, with an exception for the broader
region of the neural network. This can be related to the more
narrow degree sequence of the neural network~\cite{amaral:classes}. We
have established a correlation between $r$ and
$C$. Ref.~\cite{mejn:why} argues that such correlation occurs in
social networks because of their modularity (or ``community
structure'' as the authors call it). However, our large-$r$ networks
have no explicit bias towards high modularity, which leads us to
conjecture that the correlation between $r$ and $C$, or more
fundamentally the sum $\sum_{(i,j)\in E} k_ik_j$ (which, given a
degree sequence, is the only factor of Eq.~\ref{eq:assmix} that can
vary) is a more general phenomenon. Since $r$ is normalized by,
essentially, the variance of the degree, it follows that the valid
region for $\mathcal{G}(G)$ with more narrow degree sequence will
appear stretched (larger).

Turning to the average size of the largest component, we observe that the
gene fusion network is indeed more fragmented than the average network of the
same $(r,C)$-coordinates (as anticipated from comparing Figs.~\ref{fig:exp}(b) and (c)). The protein interaction and neural networks
have no particular bias in this respect, whereas the metabolic network
is more fragmented than expected. The relatively low $s$ of the
metabolic network can be attributed to the ``modularity'' of such
networks~\cite{zhao:meta,our:curr}. Such modules are subgraphs that
are densely connected within, and sparsely inter-connected. Sometimes
they are even disconnected from the largest component (which explains
the lower $s$). In general, $s$ decreases with assortativity. This
is natural---in more assortative networks high degree vertices are
connected to each other, forming a highly connected core and a
periphery too sparse to be connected (viz.\ Fig.~\ref{fig:exp}(c) and
(d)). For the denser networks (the protein interaction, metabolic and
neural networks) $s$ increases with $C$ (for a fixed $r$). For the
sparser gene-fusion network $s$ has a peak at
intermediate $C$. We do not speculate further about combinatorial
cause of these dependencies; but we note (comparing
e.g.\ Figs.~\ref{fig:exp}(a) and (b)) that even though the shape of
the valid regions are similar, the $s$ behavior can be qualitatively
different.

\begin{figure*}
  \centering\resizebox*{0.75\linewidth}{!}{\includegraphics{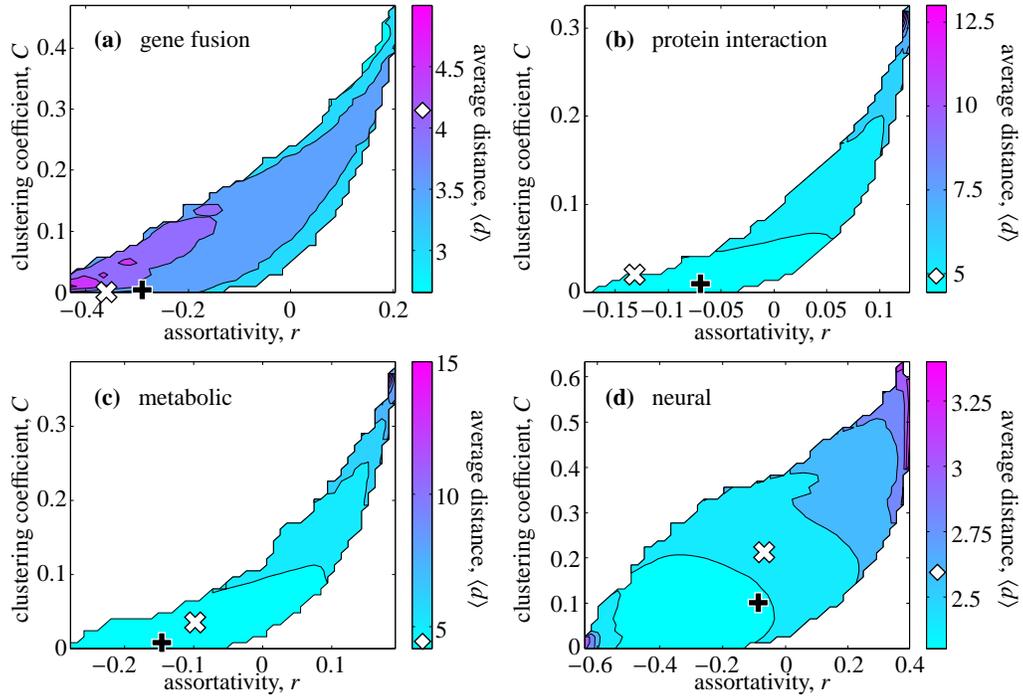}}
  \caption{The average distance within the largest component $\langle
    d\rangle$ as a function of $r$ and $C$. The panes and symbols
    correspond to those of Fig.~\ref{fig:ngc}.}
  \label{fig:length}
\end{figure*}

\subsection{Distances in the largest component}

In Fig.~\ref{fig:length} we display the average distance in the
largest component. As mentioned, measuring the distance can give complementary
information to the $s(r,C)$ graphs of Fig.~\ref{fig:ngc}---while $s$
tells us how much of the network that can be reached, $\langle
d\rangle$ tells us how fast that can happen. For all networks the big picture is that large connected
components have large average distances. This is expected from most
network models. There is, however, more information than this in
Fig.~\ref{fig:length}: for components of the same size, the average
distance is increasing with both $r$ and $C$. That $\langle d\rangle$
should increase with $C$ seems quite natural---if one of a triangle's
edges is rewired to connect two distant vertices, the distances in the
surrounding of the triangle would increase with one, but this would be
more than compensated by the connection of the two previously distant
areas. Disassortative networks typically lack a well-defined core.
Such cores are known to keep the average distance of general power-law networks
short~\cite{chung_lu:pnas}. Thus one would expect an increase of $r$
to cause a larger $\langle d\rangle$, but apparently the clustering
related length-increase outweighs this effect.
In contrast to the relative size, the average distances of the real
networks are close to the $\mathcal{G}(G)$-averages at the same
$r$-$C$ coordinates. For the gene fusion network (with a relatively
small largest component), this means the distances are rather
large.

\begin{figure*}
  \centering\resizebox*{0.75\linewidth}{!}{\includegraphics{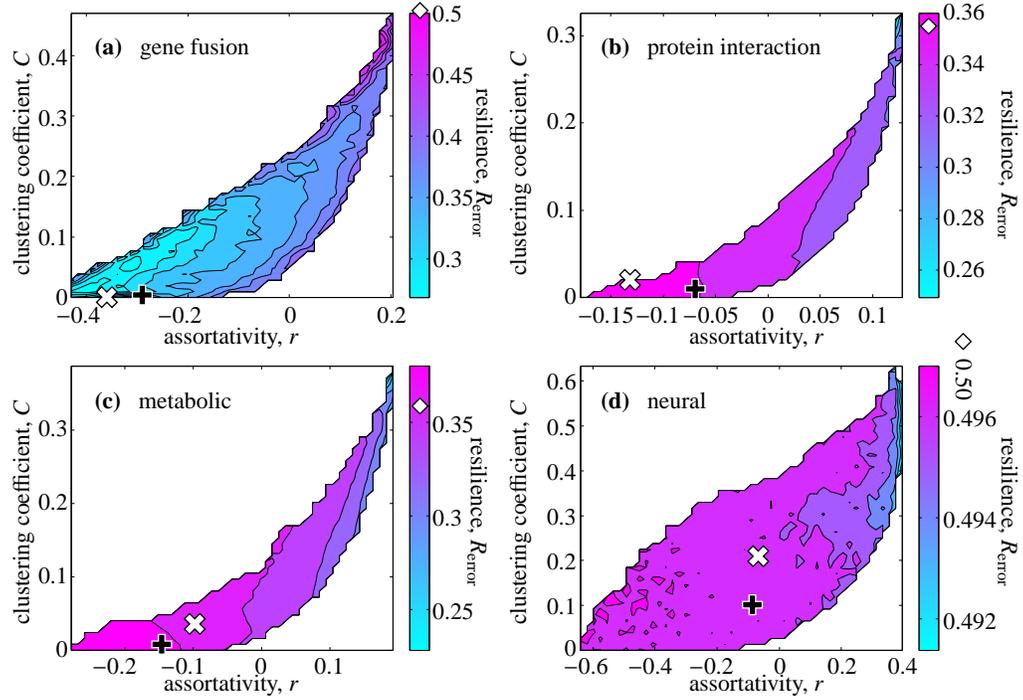}}
  \caption{The error robustness $R_\mathrm{error}$ as a function of
    $r$ and $C$. The panes and symbols correspond to those of
    Fig.~\ref{fig:ngc}.}
  \label{fig:error}
\end{figure*}

\subsection{Error robustness}

Next, we turn to the error robustness problem. As seen in
Fig.~\ref{fig:error} the gene fusion network (Fig.~\ref{fig:error}(a)),
once again, has a qualitatively different behavior than the other
three networks (Fig.~\ref{fig:error}(b), (c) and (d)). While the gene
fusion network is most robust for high $r$- and $C$-values the other
networks are most robust for low $r$. A sketchy explanation can be found
in the chain-like subgraphs extending from the largest component in
a large-$r$ network (cf.\ Fig.~\ref{fig:exp})---with a random deletion
of vertices, these subgraphs are likely to be disconnected from the
core rather soon (whereas in a disassortative network alternative paths may
still exist), then if the deletion-robust core is less than half of
the original component size it follows that it may soon be
isolated. The sparsity of the gene fusion network makes the low-$r$
$\mathcal{G}(G)$-graphs much like trees (i.e., having few cycles), and
since cycles provide redundant paths that can make a network robust,
it follows that these graphs are fragile. For a fixed
$r$, $R_\mathrm{error}$  is a decreasing function of $C$ for the three
largest networks. We believe this is an effect of the local path
redundancy induced by triangles---if one vertex of a triangle is
deleted, the other two are still connected.

The $R_\mathrm{error}$-values for the real networks are always
markedly higher than the $\mathcal{G}(G)$-averages for the same
$(r,C)$-coordinates. Networks with highly skewed degree distributions
(the gene fusion, protein interaction and metabolic networks) are
known to be robust to errors by virtue of degree distribution
alone~\cite{alb:attack}, now Fig.~\ref{fig:error} tells us that all
these networks have a yet higher error tolerance which is an indication
that error robustness is an important factor in the evolution of these
networks.

\begin{figure*}
  \centering\resizebox*{0.75\linewidth}{!}{\includegraphics{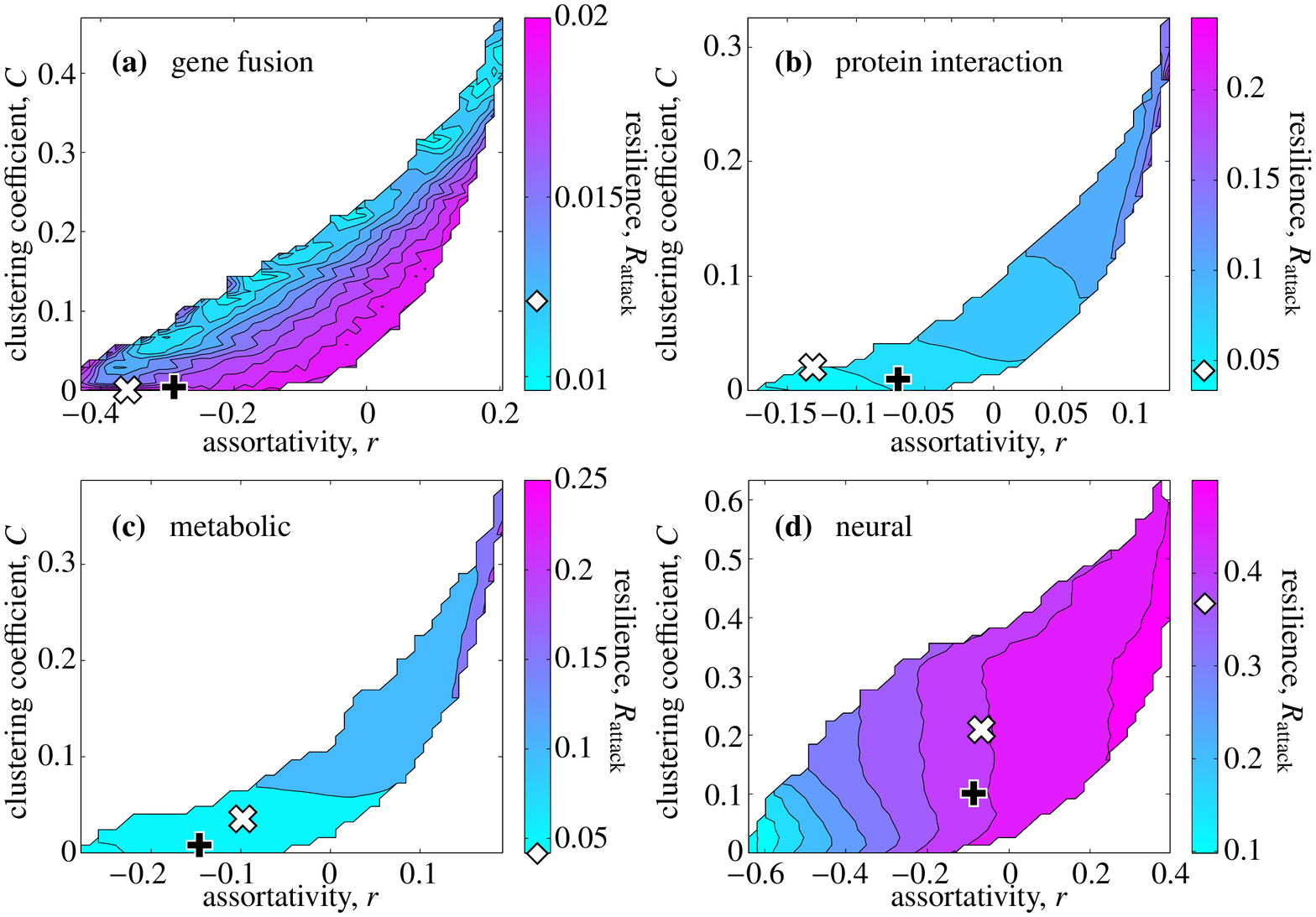}}
  \caption{The error robustness $R_\mathrm{attack}$ as a function of
    $r$ and $C$. The panes and symbols correspond to those of
    Fig.~\ref{fig:ngc}.}
  \label{fig:attack}
\end{figure*}

\subsection{Attack robustness}

The final quantity we measure is the attack robustness (see
Fig.~\ref{fig:attack}). $R_\mathrm{attack}$'s functional dependence
on $r$ and $C$ is quite different from that of $R_\mathrm{error}$. The
gene fusion $\mathcal{G}(G)$ has the highest attack robustness at high
$r$- and low $C$-values. The other networks have higher robustness
values for high assortativity, but no clear tendency in the
$C$-direction. The attack mechanism we study targets the high degree
vertices. Having all high degree vertices connected to each other is
probably the only way to keep the network from instantaneous
fragmentation. The observed $r$-dependence is thus rather
expected. The real-world networks all have $R_\mathrm{attack}$-values
of the same order of magnitude as the average values for the
$\mathcal{G}(G)$ networks of the same location in $r$-$C$ space.
We note that for studying the attack problem of metabolic networks, the (less common) enzyme centric graph representation is more appropriate (see Sect.~\ref{sect:meta}). The reason being that one can suppress an enzyme much easier than removing the substrates.

\begin{table}
\caption{Summary of the network structural measures of the real world
  networks relative to the average values of the $\mathcal{G}(G)$ a
  distance $\delta < 0.02$ from the real network. ``$<$'' indicates that
  the real network have a lower value than the corresponding
  $\mathcal{G}(G)$-value. All results are significant with p-values
  $>0.01$, except the $s$-value of the neural network that has a
  p-value of $\sim 0.05$.}
\begin{ruledtabular}
  \begin{tabular}{r|dddd}\label{tab:pval}
    & \multicolumn{1}{c}{gene fusion} &
    \multicolumn{1}{c}{protein interaction} &
    \multicolumn{1}{c}{metabolic} &
    \multicolumn{1}{c}{neural}\\\hline
    $s$ & < & < & < & > \\
    $\langle d\rangle$ & < & < & < & < \\
    $R_\mathrm{error}$ & > & > & > & > \\
    $R_\mathrm{attack}$ & > & > & > & > \\
  \end{tabular}
\end{ruledtabular}
\end{table}

\subsection{Comparison between the graphs}

Even though all our example networks are constructed from biological
data, they represent fundamentally different systems---the neural
network is spatial by nature, the protein interaction and (even more
so) the metabolic networks are the background topology for an active
dynamic system, whereas the gene fusion network is a representation
of possible but undesired events. The protein interaction, metabolic
and neural networks have one thing in common---the organism needs them
to be robust to errors (caused by injuries, mutations, disease
etc.)~\cite{wagner:robu}. As mentioned above and summarized in
Table~\ref{tab:pval} the error robustness is indeed higher for the
real networks than the $\mathcal{G}(G)$-ensemble at the same
$(r,C)$-coordinates. As mentioned above, the attack robustness of the
real network is of the same order as the $\mathcal{G}(G)$-average at
the same $(r,C)$-coordinate, but actually there is a significant
tendency that these network also are more robust to
attacks. Furthermore, the distances in the largest component, and the
relative sizes $s$ are (with the neural network $s$-value as the only
exception) smaller in the real than the $\mathcal{G}(G)$ networks.

Despite these similarities between the statistics of the
real-world networks the $r$-$C$ space of the different degree
sequences have qualitatively different network structure. Especially,
the gene fusion network behaves almost the opposite of the other
networks (at least for $s$, $\langle d\rangle$ and
$R_\mathrm{error}$). The source of this opposite behavior (as we
discuss above) is probably that it is much sparser than the other
networks. The neural network is the densest network and the only one
that do not have a power-law like degree distribution.

\section{Discussion}

Many complex network studies use the ensemble $\mathcal{G}(G)$ of
graphs with the same degree sequence as the subject graph $G$ as a
null model. In contrast to a generative network model, with a few
degrees of freedom that has to be fitted approximately, such an
ensemble has $O(N)$ degrees of freedom that can be matched exactly
with the values of $G$. We argue that $\mathcal{G}(G)$ is more than a
null model---by resolving the graphs of $\mathcal{G}(G)$ in a space
defined by some network-structural measures, one can get a picture of
the opportunities and limits there are (or has been) in the evolution
of $G$. In this work we map out $\mathcal{G}(G)$ in the
two-dimensional space defined by the clustering coefficient and the
assortativity. Then we measure other network structural quantities
throughout this space. One formal way to see our method is that we
resolve $\mathcal{G}(G)$ in the (high dimensional) space of all
sensible network measures. Then, for simplicity, we project to a few
dimensions. (The case of projection to one dimension has been studied
in a less formalized way earlier---projection to
assortativity~\cite{zj:spectrum} or a ``hierarchy''
measure~\cite{rosv:mountain}.) An interesting open question is to find
the principal components of the space of all sensible network
measures. Using four example networks from biology, we measure
average values of four network-structural quantities over the $r$-$C$
space and compare these with the values of the real networks.

The functional characteristics of the $r$-$C$ spaces varies much
between the four example networks. For example, the
\textit{C. elegans} neural network covers a much larger area of the
$r$-$C$ space, something that probably relates to its more narrow
degree distribution. The human gene fusion network, on the other hand,
has a broad degree distribution similar to the \textit{S. cerevisiae}
protein interaction and human metabolic networks, still the structural
dependency on $r$ and $C$ is very different for the gene fusion
network compared to the others. We argue that this difference stems
from the sparseness of the gene fusion network. To achieve a
comprehensive understanding about how the network structure throughout
the $r$-$C$ space depends on the degree sequence, one would need a
systematic investigation of different artificial degree sequences. In
this paper, we do not pursue this goal beyond the analysis of the four
biological data sets. The position of the real networks in the valid
region of the $r$-$C$ space adds some further information. For
example, it may have been the case that networks with lower
assortativity have been favored during the evolution of the gene fusion
network. Clustering, on the other hand, has probably not put any
constraint on the network evolution. Furthermore we compare the
network structure of the real networks with the average values of
networks in $\mathcal{G}(G)$ that are close to the $(r,C)$-coordinates
of the real network. From this analysis, we conclude that all
our four example networks are more robust to both random errors and
targeted attacks than what can be expected from a random network
constrained to the same degree distribution, assortativity and clustering
coefficient. For all networks, except maybe the gene fusion network,
this is in line with robustness being an important factor in the
network evolution. Note that in this work we assume the subject
network to be accurate. To get more valid error estimates one would
need to take the accuracy of the edges into account.

The analysis scheme presented in this paper can be further extended
and analyzed. As mentioned, it would be interesting with a quantitative
evaluation of the network-structural spaces, and how they depend on
the degree sequence. One can also try, for time-resolved data sets, to
incorporate dynamic information in the analysis by monitoring the
network-evolutionary trajectory in the $r$-$C$ space.

\acknowledgements{
  The authors thank Mikael Huss and Martin Rosvall for helpful
  suggestions and comments.
  PH acknowledges financial support from the Wenner-Gren
  Foundations and the National Science Foundation (grant
  CCR--0331580).
}

\appendix

 \begin{figure}
  \centering\resizebox*{\linewidth}{!}{\includegraphics{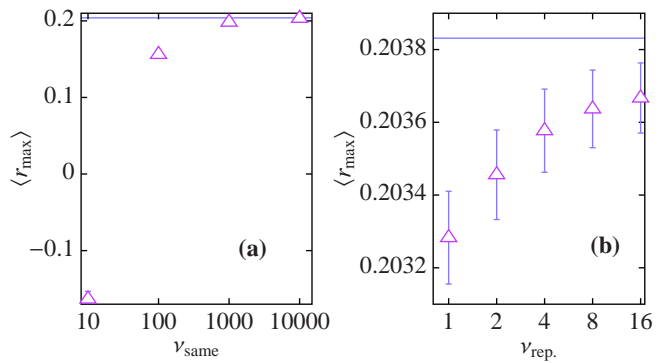}}
  \caption{Convergence of the optimization algorithm. (a) shows the
    average maximal assortativity $\langle
    r_\mathrm{max}\rangle$ with $\nu_\mathrm{rep.}=1$. The horizontal
    line represents the result of the maximization algorithm of
    Ref.~\cite{doyle:big}. (b) shows the further improvement by
    finding the maximum over many independent runs (for
    $\nu_\mathrm{same}=10000$). The vertical bars indicates the
    standard deviation of the observed maxima.}
  \label{fig:conv}
\end{figure}

\section{Convergence and sampling uniformity}

In this Appendix, we address some technical issues of our method
related to the convergence of our optimization algorithm and uniformity of
the sampling. We will also motivate our choice of parameters.

\subsection{Assortativity and clustering extremes}

To find the extremal assortativity values we use the edge-swapping
algorithm described in Sect.~\ref{sect:analysis}. To find
$r_\mathrm{min}$ we start from a random member of $\mathcal{G}(G)$ and
swap random edge-pairs (keeping the graph simple at all times) that
lower $r$. When no graph of lower $r$ has been found for
$\nu_\mathrm{same}$ time steps, we break the iteration. To avoid the
effect of being trapped in local minima, this process is repeated
$\nu_\mathrm{rep.}$ times. The main motivation for using this method
is that it is at heart the same scheme as for obtaining the extremal
clustering values and sampling the valid region (and thus we can
re-use the same code for many steps of the calculations). In this
section, we argue that the optimization performance of this method is sufficiently good for our purpose.

There is a deterministic method to maximize the assortativity that is,
if it exits properly, guaranteed to find
$r_\mathrm{max}$~\cite{doyle:big}. The method works as
follows: First all vertex-pairs $(i,j)$ are ranked in decreasing order of
the product of their degrees, $k_ik_j$. Then the edges are added
in order of this list unless the degree of one of the vertices already
is fulfilled. There are some other technicalities from the additional
constraint (of the authors) that the network should be connected. Of
our networks, only the neural network has such an evolutionary
constraint, so we do not impose it.

In Fig.~\ref{fig:conv} we display the parameter dependence of the
convergence for the gene fusion network. The horizontal line is the
theoretical maximum obtained by the algorithm of
Ref.~\cite{doyle:big}. When $\nu_\mathrm{same} = 10000$ we obtain an
average maximal assortativity within $0.001$ of the
theoretical maximum (Fig.~\ref{fig:conv}(a)). By increasing
$\nu_\mathrm{rep.}$ the accuracy can be increased further
(Fig.~\ref{fig:conv}(b)). The lattice spacing we use is $0.005\lesssim
r \lesssim 0.02$, so we deem a precision of $0.001$ sufficient. The
gene fusion network is our smallest network but the other networks are
not harder to converge. When one edge-pair is swapped so that $r$
decreases, the only term of Eq.~\ref{eq:assmix} that changes is
$\langle k_1\, k_2\rangle$. The potential change of the sum
$\sum_{(i,j)\in E} k_ik_j$, in the calculation of $\langle k_1\, k_2\rangle$
(close to the extrema) is of the order of the typical degree values
of the network. These values grow slower than the network itself,
which means that a larger network can be closer in $r$, but further
away in number of edge swaps to reach the global optimum, than a
smaller network. Some authors~\cite{doyle:big} use $\sum_{(i,j)\in E}
k_ik_j$ to measure the degree correlations, but since we strive for a
macroscopic level of description (consistent in the large-$N$ limit),
$r$ is a more appropriate quantity for the present work.

The optimization of the clustering to find the minima (maxima) of
the segments of assortativity space follows the same pattern as the
method to find the minimal (maximal) $r$. Changes of the parameters
($\nu_\mathrm{same}$ and $\nu_\mathrm{rep.}$) have the same effect as
in Fig.~\ref{fig:conv}, and the same values seem sufficient.

 \begin{figure}
  \centering\resizebox*{0.95 \linewidth}{!}{\includegraphics{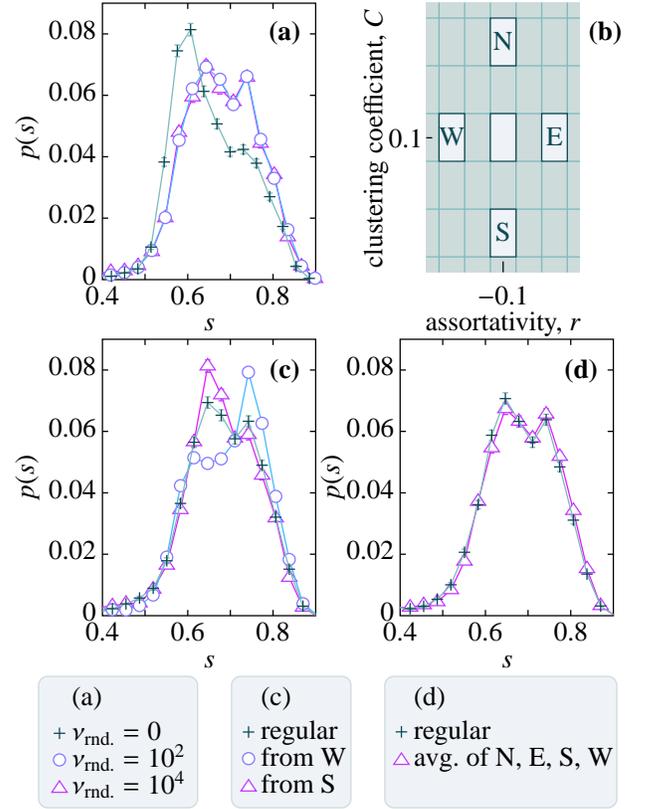}}
  \caption{Histograms of $s$ for the discussion of sampling
    uniformity. All the histograms are from the gene fusion network and a
    pixel centered around $r=-0.1$, $C=0.1$ (the dimensions of a pixel
    is $\Delta r = 0.013$, $\Delta C = 0.0096$. The error bars
    represent standard errors. Lines are guides for the eyes. (a)
    shows the histograms with a different numbers of random edge-pair
    swappings $\nu_\mathrm{rnd.}$ within the pixel before the
    measurements of quantities. (b) illustrates the location of the
    starting point pixels used in panels (c) and (d). (c) compares
    histograms for swapping processes starting at W, S with the
    regular algorithm. (d) compares the average histogram of walks starting in
    the four peripheral points of (b) with the result of the regular
    algorithm. In panels (c) and (d) $\nu_\mathrm{rnd.}=1000$. The
    whole range of the histograms is not shown, which is why the areas
    under the curves appear different.}
  \label{fig:uni}
\end{figure}

\subsection{Sampling uniformity}

The other technical issue we address in this Appendix is the
uniformity of our sampling procedure. Ideally we would like all
unique (i.e., non-isomorphic) members of $\mathcal{G}(G)$ to be
sampled with the same probability. The most important observation is
trivial---by edge-pair swapping one can go from one member of
$\mathcal{G}(G)$ to any other, and thus all members of the ensemble
will contribute to the averages. A much harder question is whether or
not every member of $\mathcal{G}(G)$ is sampled with uniform
probability. In this section, we will argue that our algorithm does a
reasonably good job in the sense that there are no inconsistencies and
parameter values are appropriate.

When the target pixel is found (step~\ref{step:rw} of the algorithm)
we perform $\nu_\mathrm{rnd.}$ additional random edge-pair swaps. The idea
is to sample the $\mathcal{G}(G)$-members of the pixel more
uniformly (and indeed to be able to reach into the interior of the
pixel). In Fig.~\ref{fig:uni}(a) we illustrate the effect of these random
moves. We plot a normalized histogram of the relative largest cluster
size $s$ for $0$, $100$ and $10000$ random moves. We see that these
moves do make a difference (the $\nu_\mathrm{rnd.}=0$ is different
from the $\nu_\mathrm{rnd.}=100$) but it does not matter if
$\nu_\mathrm{rnd.}=100$ or $\nu_\mathrm{rnd.}=10000$. The same
situation is observed for other pixels, networks and
quantities. Therefore, we use $\nu_\mathrm{rnd.}=1000$ in this work.

Next, we will illustrate the use of the randomly permuted list
in the sampling of the pixels (steps~\ref{step:perm} and
\ref{step:pick} of the algorithm). The motivation for this procedure
is that the network structure can depend on the direction from which
the search arrives to the pixel. In Fig.~\ref{fig:uni}(b) we
illustrate the test procedure---we sample separate histograms from four
starting points in the four cardinal directions with respect to the central
$(r,C)=(-0.1,0.1)$ pixel. In Fig.~\ref{fig:uni}(c) we
see that the histograms from the W and S pixels are different. There
appears to be two regions of $\mathcal{G}(G)$ contributing to these
histograms (one with $s\approx 0.65$, one with $s\approx
0.75$). Searches starting from W seem to arrive at the $s\approx 0.75$
region more frequently, and searches staring at S ends up around
$s\approx 0.65$ more frequently. The curve of the actual algorithm
weighs the two peaks more equal. The curves from N and E coincides
almost completely the curve for the regular algorithm (and are
therefore omitted for clarity). The impression we get is that the
search from one direction can induce a bias in the network structure
(symbolically speaking, the graphs have a preference for ending up in
a certain region of $\mathcal{G}(G)$). However, from other directions,
or by the random sampling of pixels (step~\ref{step:perm}), the bias is
reduced. This picture is further strengthened in Fig.~\ref{fig:uni}(d)
where we show that the average value of the histograms from the four
starting points are overlapping with the histogram of the regular
algorithm.

\end{document}